\documentclass[12pt]{article}
\usepackage[latin9]{inputenc}
\usepackage{amsmath}
\usepackage{amssymb}
\usepackage[unicode=true,pdfusetitle,
 bookmarks=true,bookmarksnumbered=false,bookmarksopen=false,
 breaklinks=false,pdfborder={0 0 0},pdfborderstyle={},backref=false,colorlinks=false]
 {hyperref}

\makeatletter

\DeclareTextSymbolDefault{\textquotedbl}{T1}

\numberwithin{equation}{section}

\usepackage{amsfonts}
\usepackage{ytableau} 
\usepackage[square,numbers,sort&compress]{natbib}

\makeatother

\begin{document}
\begin{flushright}
FIAN/TD/4-2024
\par\end{flushright}

\vspace{0.5cm}
 
\begin{center}
\textbf{\large{}Scalar Electrodynamics and Higgs Mechanism}\\
\textbf{\large{}in the Unfolded Dynamics Approach}{\large\par}
\par\end{center}

\begin{center}
\par\end{center}

\begin{center}
\vspace{0.2cm}
 \textbf{Nikita~Misuna}\\
 \vspace{0.5cm}
 \textit{Tamm Department of Theoretical Physics, Lebedev Physical
Institute,}\\
 \textit{Leninsky prospekt 53, 119991, Moscow, Russia}\\
 
\par\end{center}

\begin{center}
\vspace{0.6cm}
 misuna@lpi.ru \\
 
\par\end{center}

\vspace{0.4cm}

\begin{abstract}
\noindent We put forward a novel method of constructing unfolded formulations
of field theories, which is based on initial fixation of the form
of an unfolded field and subsequent looking for the corresponding
unfolded equation as an identity that this field satisfies. Making
use of this method, we find an unfolded formulation for 4d scalar
electrodynamics. By considering a symmetry-breaking scalar potential,
we study the implementation of the Higgs mechanism within the framework
of the unfolded dynamics approach. We explore a deformation of unfolded
modules in the symmetry-broken phase and identify a non-invertible
unfolded-field redefinition that diagonalizes the higgsed system.

\newpage{}

\tableofcontents{}
\end{abstract}

\section{Introduction}

Symmetries play the central role in modern theories of fundamental
interactions. One of the most complete implementations of this principle
is given by higher-spin (HS) gravity theories (for a partial review
of the recent literature on the topic see \cite{snow}). These are
theories containing interacting massless fields of all spins, that
leads to the emergence of infinite-dimensional HS gauge symmetry \cite{HSalgebra}.
This makes HS gravities promising candidates for the role of quantum
gravity theory. It is also suggested that HS gravities may represent
a symmetric \textquotedbl Coulombian\textquotedbl{} phase of string
theory at trans-Planckian energies \cite{string1,string2,string3}.

In order to keep HS symmetry manifest, a special formalism for operating
with HS gravity has been developed, called unfolded dynamics approach
\cite{unf1,vas1,vas2,unf2,ActionsCharges}. Within the unfolded framework,
a field theory is formulated as a set of first-order differential
equations on unfolded fields, being exterior forms. These unfolded
fields encode all d.o.f. of the theory, so usually a spectrum of unfolded
fields is infinite or, equivalently, unfolded fields are defined in
some larger space, equipped with additional coordinates besides space-time
ones. This is the price to pay for having a coordinate-independent
manifestly gauge-invariant first-order formulation.

It is of natural interest to try to apply this formalism to various
models beyond HS gravities of \cite{vas1,vas2,did1,did2}. Up to now,
very few such unfolded nonlinear theories are available \cite{unfQFT,ActionsCharges,unf_conf}.
The reason behind this is that a general consistency analysis, which
is a standard tool for unfolding, becomes drastically more complicated
in the nonlinear case.

In this paper we put forward a novel method of constructing unfolded
formulations for theories, which is based on postulating a concrete
form of an unfolded field and further looking for corresponding unfolded
equations as for identities satisfied by this field. We successfully
apply this method to $4d$ scalar electrodynamics, obtaining its unfolded
formulation. We then study the spontaneous symmetry breaking in this
theory within the unfolded framework, which is of particular interest
in light of recent studies of symmetry breaking in HS gravity \cite{HSbreak}.

The paper is organized as follows. In Section \ref{SEC_UDA}, we give
a flash review of the unfolded dynamics approach and present a new
method of unfolding with the example of a $4d$ self-interacting scalar
model. In Section \ref{SEC_QED}, we make use of this method in order
to construct an unfolded formulation of $4d$ scalar electrodynamics.
Then in Section \ref{SEC_Higgs}, we analyze Higgs mechanism in the
unfolded system we built. In Section \ref{SEC_conclusions}, we present
our conclusions.

\section{Unfolded Dynamics Approach\label{SEC_UDA}}

In this Section, we briefly discuss a general construction of the
unfolded dynamics approach and consider two relevant examples: an
unfolded non-dynamical Minkowski background and an unfolded self-interacting
scalar field.

\subsection{General construction\label{SUB_General_construction}}

Unfolded dynamics approach \cite{unf1,vas1,vas2,unf2,ActionsCharges}
to a field theory consists in representing it in the form of \textquotedbl unfolded\textquotedbl{}
first-order equations
\begin{equation}
\mathrm{d}W^{A}(x)+G^{A}(W)=0,\label{unf_eq}
\end{equation}
where$\mathrm{d}$ is the exterior derivative on a space-time manifold
$M^{d}$ and unfolded fields $W^{A}(x)$ are exterior forms on $M^{d}$,
with $A$ standing for all indices of the field. $G^{A}(W)$ is built
from exterior products of unfolded fields (the wedge symbol is omitted
throughout the paper). There is one and only one unfolded equation
\eqref{unf_eq} for every unfolded field $W^{A}$.

The identity $\mathrm{d}^{2}\equiv0$ imposes a consistency condition
on $G$
\begin{equation}
G^{B}\dfrac{\delta G^{A}}{\delta W^{B}}\equiv0.\label{unf_consist}
\end{equation}

Unfolded equations \eqref{unf_eq} are manifestly invariant under
infinitesimal gauge transformations 
\begin{equation}
\delta W^{A}=\mathrm{d}\varepsilon^{A}(x)-\varepsilon^{B}\dfrac{\delta G^{A}}{\delta W^{B}}\label{unf_gauge_transf}
\end{equation}
with a gauge parameter $\varepsilon^{A}(x)$ being a $(n-1)$-form
for a $n$-form $W^{A}$. A spectrum of unfolded fields is usually
infinite, because it corresponds to all d.o.f. of the theory. Typically,
there is some grading bounded from below on the space of unfolded
fields. Then equations \eqref{unf_eq} relate higher-grade fields
to the space-time derivatives of the lower-grade ones and, at the
same time, impose dynamical constraints on the lowest-grade fields.
For this reason, lowest-grade unfolded fields are referred to as primary
fields, while the higher-grade ones are referred to as descendants.

An unfolded formulation provides a manifestly coordinate-independent
and gauge-invariant description of a theory. Both of these features
are of critical importance for HS gravity. The first-order nature
of the formalism can potentially help in studying integrability (in
particular, the problem of looking for conserved charges within the
unfolded framework becomes the cohomology problem for some operator
determined by \eqref{unf_eq} \cite{ActionsCharges}). All this makes
it prominent to apply the unfolded dynamics approach to the field
theories beyond the realm of HS gravity. However, constructing unfolded
formulations (especially for nonlinear theories) is not an easy task.
In this paper we put forward a novel method of unfolding, which allows
us to construct an unfolded formulation for scalar electrodynamics
with an arbitrary scalar potential. But first, in this Section we
demonstrate this method using the example of a self-interacting scalar
theory.

\subsection{Unfolded Minkowski background}

The background geometry of $M^{d}$ is expressed via imposing the
Maurer\textendash Cartan equation on a 1-form connection $\Omega$
taking values in the Lie algebra of symmetries of $M^{d}$
\begin{equation}
\mathrm{d}\Omega+\frac{1}{2}[\Omega,\Omega]=0,\label{flat_conn}
\end{equation}
where square brackets stand for the Lie-algebra commutator. Global
symmetries of $M^{d}$ arise as a residual symmetry \eqref{unf_gauge_transf},
which is left over after choosing some particular $\Omega_{0}$ that
solves \eqref{flat_conn},
\begin{equation}
\mathrm{d}\varepsilon_{0}+[\Omega_{0},\varepsilon_{0}]=0.\label{glob_symm}
\end{equation}
In this paper, we deal with $4d$ Minkowski space, so the Lie algebra
in question is $iso(1,3)$ and the corresponding connection is
\begin{equation}
\Omega=e^{\alpha\dot{\beta}}P_{\alpha\dot{\beta}}+\omega^{\alpha\beta}M_{\alpha\beta}+\bar{\omega}^{\dot{\alpha}\dot{\beta}}\bar{M}_{\dot{\alpha}\dot{\beta}},\label{Poincare_connection}
\end{equation}
with $P_{\alpha\dot{\alpha}}$, $M_{\alpha\beta}$ and $\bar{M}_{\dot{\alpha}\dot{\beta}}$
being generators of translations and rotations of $\mathbb{R}^{1,3}$,
$e^{\alpha\dot{\beta}}$ and $\omega^{\alpha\beta}$ ($\bar{\omega}^{\dot{\alpha}\dot{\beta}}$)
being 1-forms of vierbein and Lorentz connection. Greek indices correspond
to two-component (Weyl) spinor representations. They are moved by
an antisymmetric Lorentz-invariant spinor metric
\begin{equation}
\epsilon_{\alpha\beta}=\epsilon_{\dot{\alpha}\dot{\beta}}=\left(\begin{array}{cc}
0 & 1\\
-1 & 0
\end{array}\right),\quad\epsilon^{\alpha\beta}=\epsilon^{\dot{\alpha}\dot{\beta}}=\left(\begin{array}{cc}
0 & 1\\
-1 & 0
\end{array}\right)
\end{equation}
as
\begin{equation}
v_{\alpha}=\epsilon_{\beta\alpha}v^{\beta},\quad v^{\alpha}=\epsilon^{\alpha\beta}v_{\beta},\quad\bar{v}_{\dot{\alpha}}=\epsilon_{\dot{\beta}\dot{\alpha}}\bar{v}^{\dot{\beta}},\quad\bar{v}^{\dot{\alpha}}=\epsilon^{\dot{\alpha}\dot{\beta}}\bar{v}_{\dot{\beta}}.
\end{equation}

The simplest (non-degenerate) solution to \eqref{flat_conn} with
$\Omega$ being \eqref{Poincare_connection} is provided by global
Cartesian coordinates
\begin{equation}
e_{\underline{m}}{}^{\alpha\dot{\beta}}=(\bar{\sigma}_{\underline{m}})^{\dot{\beta}\alpha},\quad\omega_{\underline{m}}{}^{\alpha\beta}=0,\quad\bar{\omega}_{\underline{m}}{}^{\dot{\alpha}\dot{\beta}}=0.\label{cartes_coord}
\end{equation}
Then the general solution to \eqref{glob_symm} determines 10 parameters
of global Poincaré transformations.

Here the spectrum of unfolded field is finite, containing only the
1-form $\Omega$, because this system is non-dynamical.

\subsection{Example: unfolded self-interacting scalar\label{SUB_NLIN_SCALAR}}

Now let us consider two unfolding procedures for the theory of a $4d$
self-interacting scalar field: the standard one, implemented in \cite{unfQFT},
and a novel one, which is simpler and more convenient, allowing the
application to scalar electrodynamics.

A standard strategy is based on studying the consistency condition
\eqref{unf_consist}: one assumes some spectrum of unfolded fields,
then writes down an appropriate ansatz for unfolded equations \eqref{unf_eq}
and finally tries to fix it by solving for the consistency condition
\eqref{unf_consist} (which, in its turn, may force one to modify
the initially assumed field spectrum and, accordingly, the ansatz).
This can be performed for linear theories (see e.g. \cite{unf3,unf4,unf5,unf6,unf7,unf8,unf9,unf10}),
but for nonlinear models this method is not particularly productive.
The consistency equation is of higher order in fields than the initial
ansatz, so one ends up with a complicated system of entangled equations
on coefficients in the ansatz. Thus, in general it is uneasy to constrain
sufficiently the form of the ansatz so that the procedure is practicable.

As an example, consider unfolding a nonlinear scalar theory
\begin{equation}
(\square+m^{2})\phi+\mathrm{U}'(\phi)=0,\label{nlin_KG}
\end{equation}
where $\mathrm{U}'$ is the variation of the scalar potential.

This has been unfolded in \cite{unfQFT} by analyzing consistency
(in fact, a more general off-shell case has been solved there, but
that does not interest us here). There an ansatz was guessed, that
was simple enough to allow for direct analysis. Let us sketch that
derivation.

A spectrum of unfolded fields represents a set of 0-forms that can
be collected into a single unfolded master-field
\begin{equation}
\Phi(Y|x)=\sum_{n=0}^{\infty}\Phi_{\alpha(n),\dot{\alpha}(n)}(x)y^{\alpha_{1}}...y^{\alpha_{n}}\bar{y}^{\dot{\alpha}_{1}}...\bar{y}^{\dot{\alpha}_{n}},\label{F_scalar}
\end{equation}
where the condensed index notations are used
\begin{equation}
f_{\alpha(n)}:=f_{\alpha_{1}...\alpha_{n}},
\end{equation}
and a pair of auxiliary commuting Weyl spinors $Y=(y^{\alpha},\bar{y}^{\dot{\alpha}})$
is introduced for the convenience of operating with symmetric spinor-tensors.
Due to their commutativity, $Y$ are null with respect to the spinor
metric 
\begin{equation}
y^{\alpha}y^{\beta}\epsilon_{\alpha\beta}=0,\quad\bar{y}^{\dot{\alpha}}\bar{y}^{\dot{\beta}}\epsilon_{\dot{\alpha}\dot{\beta}}=0.\label{null_Y}
\end{equation}
The master-field \eqref{F_scalar} corresponds to an infinite set
of symmetric traceless Lorentz tensors of all ranks, as can be seen
from contracting all spinor indices of $\Phi_{\alpha(n),\dot{\alpha}(n)}$
with $\sigma$-matrices 
\begin{equation}
\Phi_{a_{1}a_{2}...a_{n}}=(\bar{\sigma}_{a_{1}})^{\dot{\alpha}_{1}\alpha_{1}}...(\bar{\sigma}_{a_{n}})^{\dot{\alpha}_{n}\alpha_{n}}\Phi_{\alpha(n),\dot{\alpha}(n)},\quad\eta^{a_{1}a_{2}}\Phi_{a_{1}a_{2}...a_{n}}=0.
\end{equation}
This is the unfolded spectrum of a free scalar field \cite{unf3}
and a scalar field of HS gravity \cite{unf1}, so it is natural to
take it for the problem in question.

We also introduce spinorial derivatives
\begin{equation}
\partial_{\alpha}y^{\beta}=\delta_{\alpha}\text{}^{\beta},\quad\bar{\partial}_{\dot{\alpha}}\bar{y}^{\dot{\beta}}=\delta_{\dot{\alpha}}\text{}^{\dot{\beta}}
\end{equation}
and spinorial Euler operators
\begin{equation}
N=y^{\alpha}\partial_{\alpha},\quad\bar{N}=\frac{1}{2}\bar{y}^{\dot{\alpha}}\bar{\partial}_{\dot{\alpha}}.\label{Euler}
\end{equation}
For a scalar master-field \eqref{nlin_KG} two Euler operators in
fact coincide
\begin{equation}
N\Phi=\bar{N}\Phi.
\end{equation}

The Euler operator can be taken as a grading operator discussed in
Subsection \ref{SUB_General_construction}. We will see that $Y$-dependent
components of the master-field \eqref{F_scalar} are differential
descendants of the primary scalar field, which is a $Y$-independent
component
\begin{equation}
\phi(x)=\Phi(Y=0|x).\label{primary_phi}
\end{equation}

In \cite{unfQFT}, the following ansatz for an unfolded equation was
proposed
\begin{equation}
\mathrm{d}_{L}\Phi-a_{N}e\partial\bar{\partial}\Phi+b_{N}ey\bar{y}m^{2}\Phi+c_{N}ey\bar{y}\mathrm{U}'(f_{N}\Phi)=0,\label{scalar_ansatz}
\end{equation}
where
\begin{equation}
e\partial\bar{\partial}:=e^{\alpha\dot{\beta}}\partial_{\alpha}\bar{\partial}_{\dot{\beta}},\quad ey\bar{y}:=e^{\alpha\dot{\beta}}y_{\alpha}\bar{y}_{\dot{\beta}},
\end{equation}
coefficients $a_{N}$, $b_{N}$, $c_{N}$, $f_{N}$ depend on Euler
operator $N$ \eqref{Euler}, every $f_{N}$ in $\mathrm{U}'$ acts
on a single factor of $\Phi$ and $\mathrm{d}_{L}$ is the Lorentz-covariant
derivative
\begin{equation}
\mathrm{d}_{L}f(Y|x):=\left(\mathrm{d}+\omega^{\alpha\beta}y_{\alpha}\partial_{\beta}+\bar{\omega}^{\dot{\alpha}\dot{\beta}}\bar{y}_{\dot{\alpha}}\bar{\partial}_{\dot{\beta}}\right)f(Y|x),
\end{equation}
which in Cartesian coordinates \eqref{cartes_coord} comes down to
the exterior derivative. 

A general solution (up to constant rescaling of $\Phi$, $m^{2}$
and $\mathrm{U}$) to the corresponding consistency condition \eqref{unf_consist}
for the equation \eqref{scalar_ansatz} can be formulated in terms
of dependence of coefficients on arbitrary (but necessarily non-zero)
$a_{N}$
\begin{equation}
b_{N}=\frac{1}{N(N+1)a_{N-1}},\quad c_{N}=\frac{1}{(N+1)!\prod_{i=0}^{N-1}a_{i}},\quad f_{N}=N!\prod_{i=0}^{N-1}a_{i}.\label{coeff_sol}
\end{equation}
Then $Y$-dependence of $\Phi$ is manifestly resolved as 
\begin{equation}
\Phi(Y|x)=\sum_{n=0}^{\infty}\frac{(y^{\alpha}\bar{y}^{\dot{\alpha}}\nabla_{\alpha\dot{\alpha}})^{n}}{(n!)^{2}\prod_{i=0}^{n-1}a_{i}}\phi(x),\label{scalar_Y_dependence}
\end{equation}
where a 0-form derivative $\nabla_{\alpha\dot{\alpha}}$ is introduced
via 
\begin{equation}
\mathrm{d}_{L}=e^{\alpha\dot{\alpha}}\nabla_{\alpha\dot{\alpha}}\label{nabla_def}
\end{equation}
(in Cartesian coordinates, $\nabla_{\alpha\dot{\alpha}}$ comes down
to the usual partial derivative) and the primary scalar $\phi(x)$
is subjected to the nonlinear Klein\textendash Gordon equation \eqref{nlin_KG}
with
\begin{equation}
\square:=\frac{1}{2}\nabla_{\alpha\dot{\alpha}}\nabla^{\alpha\dot{\alpha}}.
\end{equation}
For the sake of simplicity, we assume that $\nabla_{\alpha\dot{\alpha}}$
commutes with $Y$, because one can always choose Cartesian coordinates.
Then the covariance of the final formulas can be achieved by supplementing
all partial $x$-derivatives with $\omega$-terms, since this is the
only way the Lorentz-connection can enter the equations.

The dependence \eqref{scalar_Y_dependence} can be found by acting
with 
\begin{equation}
y^{\alpha}\bar{y}^{\dot{\beta}}\frac{\delta}{\delta e^{\alpha\dot{\beta}}}
\end{equation}
on \eqref{scalar_ansatz}. Then \eqref{nlin_KG} result from acting
on \eqref{scalar_ansatz} with
\begin{equation}
(\nabla^{\alpha\dot{\alpha}}+a_{N}\partial^{\alpha}\bar{\partial}^{\dot{\alpha}})\frac{\delta}{\delta e^{\alpha\dot{\alpha}}}.
\end{equation}

Thus, the unfolded system \eqref{scalar_ansatz} with coefficients
obeying \eqref{coeff_sol} indeed describes the theory of a self-interacting
scalar. $Y$-dependent components of the unfolded master-field $\Phi$
\eqref{F_scalar} represent descendants (traceless space-time derivatives)
of the primary (with respect to the $N$-grading) scalar field $\phi(x)$,
subjected to the nonlinear Klein\textendash Gordon equation \eqref{nlin_KG}.

In this paper, we put forward a different method of unfolding a field
theory: one should first postulate some concrete form of an unfolded
master-field and then look for a corresponding unfolded equation,
identically satisfied by this master-field. This last step in practice
consists of expressing an action of $e^{\alpha\dot{\beta}}\partial_{\alpha}\bar{\partial}_{\dot{\beta}}$
on the master-field in terms of $\mathrm{d}_{L}$ and other spinor
operators acting on unfolded fields. Let us demonstrate how this works
for the problem of unfolding the nonlinear scalar.

We start by postulating that an unfolded scalar master-field is
\begin{equation}
\Phi(Y|x)=e^{y^{\alpha}\bar{y}^{\dot{\alpha}}\nabla_{\alpha\dot{\alpha}}}\phi(x),\label{F_postulate}
\end{equation}
with the primary field $\phi(x)$ subjected to \eqref{nlin_KG}. To
simplify the appearance of formulas, we further omit spinor indices
contracted between spinors and the derivative
\begin{equation}
\nabla y\bar{y}:=y^{\alpha}\bar{y}^{\dot{\alpha}}\nabla_{\alpha\dot{\alpha}}.
\end{equation}

Now we need to find an appropriate unfolded equation, whose solution
is \eqref{F_postulate}. To this end, we use the following identity,
which holds for an arbitrary function $f$,
\begin{equation}
\partial_{\alpha}\bar{\partial}_{\dot{\alpha}}f(z_{\beta\dot{\beta}}y^{\beta}\bar{y}^{\dot{\beta}})=z_{\alpha\dot{\alpha}}(N+1)f'-\frac{1}{2}y_{\alpha}\bar{y}_{\dot{\alpha}}z_{\beta\dot{\beta}}z^{\beta\dot{\beta}}f'',
\end{equation}
where the prime denotes a derivative of $f$ with respect to its full
argument. Then we have
\begin{equation}
\partial_{\alpha}\bar{\partial}_{\dot{\alpha}}\Phi(Y|x)=(N+1)\nabla_{\alpha\dot{\alpha}}e^{\nabla y\bar{y}}\phi(x)-y_{\alpha}\bar{y}_{\dot{\alpha}}\square e^{\nabla y\bar{y}}\phi(x),\label{ddPhi_non_gauge}
\end{equation}
and hence
\begin{equation}
\partial_{\alpha}\bar{\partial}_{\dot{\alpha}}\Phi=(N+1)\nabla_{\alpha\dot{\alpha}}\Phi+y_{\alpha}\bar{y}_{\dot{\alpha}}(m^{2}\Phi+\mathrm{U}'(\Phi)).
\end{equation}
Contracting this with the vierbein, we get the desired unfolded equation
\begin{equation}
\mathrm{d}_{L}\Phi-\frac{1}{N+1}e\partial\bar{\partial}\Phi+\frac{1}{N+1}ey\bar{y}\left(m^{2}\Phi+\mathrm{U}'(\Phi)\right)=0,\label{F_eq}
\end{equation}
which is the particular case of the general solution \eqref{coeff_sol},
corresponding to the choice
\begin{equation}
a_{N}=\frac{1}{N+1}.
\end{equation}

Note that in our analysis the resulting unfolded equation \eqref{F_eq}
is consistent by construction, because it arises as the identity satisfied
by the unfolded field \eqref{F_postulate} that we started with. This
also means that this equation, being nonlinear (the form of the potential
$\mathrm{U}(\phi)$ is unrestricted), is manifestly integrable (in
the sense of restoring the $Y$-dependence) by construction, as we
know its solution. This, in its turn, allows us to generate all other
particular unfolded systems \eqref{scalar_ansatz} and/or immediately
obtain their solutions.

Let us illustrate the last claim. Suppose one wants to generate an
unfolded system of the type \eqref{scalar_ansatz} with some given
$a_{N}$. Having the equation \eqref{F_eq} with its solution \eqref{F_postulate}
in hand, there is no need to repeat the analysis again. One just needs
to redefine the unfolded master-field with a for-now arbitrary coefficient
$\rho_{N}$
\begin{equation}
\Phi(Y|x)=\rho_{N}\widetilde{\Phi}(Y|x).\label{F_tilded}
\end{equation}
Substituting this into \eqref{F_eq} yields
\begin{equation}
\mathrm{d}_{L}\widetilde{\Phi}-\frac{\rho_{N+1}}{\rho_{N}(N+1)}e\partial\bar{\partial}\widetilde{\Phi}+\frac{\rho_{N-1}}{\rho_{N}(N+1)}ey\bar{y}m^{2}\widetilde{\Phi}+\frac{1}{\rho_{N}(N+1)}ey\bar{y}\mathrm{U}'(\rho_{N}\widetilde{\Phi})=0.\label{F_tilded_eq}
\end{equation}
Now, demanding
\begin{equation}
\frac{\rho_{N+1}}{\rho_{N}(N+1)}=a_{N},
\end{equation}
one finds
\begin{equation}
\rho_{N}=\rho_{0}\cdot N!\prod_{i=0}^{N-1}a_{i}\label{rho_sol}
\end{equation}
with arbitrary (non-zero) $\rho_{0}$. 

Thus, one gets a consistent unfolded system of the required form \eqref{F_tilded_eq}
with coefficients determined by \eqref{rho_sol} and with the solution
\begin{equation}
\widetilde{\Phi}(Y|x)=\frac{1}{\rho_{N}}e^{\nabla y\bar{y}}\phi(x),\label{F_tilded_sol}
\end{equation}
as follows from \eqref{F_tilded}. For example, if one takes $a_{N}=1$,
which is a common choice in the HS literature, this gives
\begin{equation}
\mathrm{d}_{L}\widetilde{\Phi}-e\partial\bar{\partial}\widetilde{\Phi}+\frac{1}{N(N+1)}ey\bar{y}m^{2}\widetilde{\Phi}+\frac{1}{(N+1)!}ey\bar{y}\mathrm{U}'(N!\widetilde{\Phi})=0,\label{F_eq_on_shell-1}
\end{equation}
\begin{equation}
\widetilde{\Phi}(Y|x)=\sum_{n=0}^{\infty}\frac{(\nabla y\bar{y})^{n}}{(n!)^{2}}\phi(x)={}_{0}F_{1}(;1;\nabla y\bar{y})\phi(x),
\end{equation}
so in this case an unfolding operator is a confluent hypergeometric
limit function \cite{unfQFT}.

This also allows one to immediately write down an unfolded system
for a given form of the master-field. To this end one just needs to
use \eqref{F_tilded_sol} in order to find corresponding $\rho_{N}$,
then the required equation is \eqref{F_tilded_eq}. Suppose that one
needs an unfolded system that leads to the master-field of the form
\begin{equation}
\widetilde{\Phi}(Y|x)=\sum_{n=0}^{\infty}k_{n}(\nabla y\bar{y})^{n}\phi(x)\label{Phi_general}
\end{equation}
with all $k_{n}$ being non-zero. Comparing with \eqref{F_tilded_sol},
one finds corresponding $\rho_{N}$ to be
\begin{equation}
\rho_{N}=\frac{1}{k_{N}\cdot N!}.
\end{equation}

Thus, although the proposed method is aimed at constructing a particular
unfolded formulation, in effect it allows one to easily reproduce
all the results of the general consistency analysis. For a nonlinear
gauge theory considered in the paper, the consistency analysis is
in fact impracticable, so the new method remains the only available
tool.

Let us also note that, although we were considering the nonlinear
theory, all unfoldings \eqref{scalar_Y_dependence} and \eqref{F_postulate}
were linear. It means that, in general, interactions do not deform
the unfolding map. In fact, it is \emph{gauge} interactions that do.
Scalar electrodynamics, considered in the next Section, provides an
example. But even here one can see that introducing dynamical gravity
would make the unfolding nonlinear, since the operator $\nabla_{\alpha\dot{\alpha}}$
is defined in terms of the vierbein \eqref{nabla_def}.

Summing up all these observations, we can determine the level of generality
of the ansatz \eqref{scalar_ansatz}. We see that it covers all unfoldings
$\phi(x)\rightarrow\Phi(Y|x)$ of the form \eqref{Phi_general}, i.e.
all unfoldings which are regular and linear. Let us discuss both of
these properties.

The form of the master-field \eqref{Phi_general} implies that the
primary $\phi(x)$ is infinitely differentiable (at least up to d'Alembertians),
otherwise the master-field does not exist. However, the unfolded equation
\eqref{scalar_ansatz} is of first order and local in $x$ and thus
allows for singular solutions, which means that the representation
\eqref{Phi_general} may be valid only locally. Solutions which are
singular in $x$ and/or in $Y$ are important in HS gravity, in particular
in the context of HS black holes \cite{bh1,bh2,bh3,bh4,bh5}.

Another point is that one may consider nonlinear unfoldings like e.g.
\begin{equation}
\Phi^{nonlin}(Y|x)=\sum_{n=0}^{\infty}k_{n}(\nabla y\bar{y})^{n}\phi(x)+\sum_{n=1}^{\infty}\ell_{n}(\nabla y\bar{y})^{n}\phi^{2}(x),
\end{equation}
or more complicated ones. They still have the same primary component
$\Phi^{nonlin}(Y=0|x)=\phi(x)$ (this is why the second sum starts
with $n=1$), but the relation between $Y$-variables and $x$-derivatives
of $\phi(x)$ is highly sophisticated now. So the method set out here
is not directly applicable to such unfoldings. Perhaps, this may serve
as a general guiding principle: one should look for unfoldings which
are linear in non-gauge fields (even for interacting theories), since
they allow for a direct local space-time interpretation of auxiliary
$Y$-variables.

Finally, let us mention that in principle there is a different route
to unfold dynamical nonlinear theories. Namely, one can generate corresponding
unfolded equations by quotienting the space of off-shell (i.e. non-dynamical)
unfolded fields by an invariant subspace, spanned by the differential
descendants that are put to zero by dynamical equations. Let us illustrate
this by deriving a dynamical nonlinear scalar field theory from a
linear off-shell one.

We consider an off-shell unfolded master-field
\begin{equation}
\Psi(Y,\tau|x)=e^{\tau\square+y\bar{y}\nabla}\phi(x),\label{F_off-shell_postulate}
\end{equation}
with the primary field $\phi(x)$ completely unconstrained. A new
auxiliary scalar variable $\tau$ encodes an expansion in d'Alembertians
of $\phi(x)$, which are now unfixed \cite{unf8}. The corresponding
unfolded equation, which can be deduced analogously to the dynamical
case considered above, is \cite{unfQFT}
\begin{equation}
\mathrm{d}_{L}\Psi-\frac{1}{N+1}e\partial\bar{\partial}\Psi-\frac{1}{N+1}ey\bar{y}\frac{\partial}{\partial\tau}\Psi=0.\label{F_off-shell_eq}
\end{equation}
This unfolded system just encodes an infinite set of constraints that
express all descendants in terms of the primary $\phi(x)$ leaving
it unconstrained. So, in a sense, all off-shell systems with the same
spectrum of primary fields are formally equivalent, differing only
in the specific way in which descendants are parameterized (which,
however, can affect such important points as regularity, locality
etc.).

Now we want to impose a nonlinear dynamical equation \eqref{nlin_KG}
on $\phi(x)$ . Considering \eqref{F_off-shell_postulate}, we see
that it is equivalent to the following nonlinear constraint on $\Psi$
\begin{equation}
(\frac{\partial}{\partial\tau}\Psi+m^{2}\Psi+\mathrm{U}'(\Psi))|_{\tau=0}=0.\label{off-shell_quotient}
\end{equation}
Thus, the linear off-shell unfolded system \eqref{F_off-shell_eq}
endowed with the nonlinear \textquotedbl initial condition\textquotedbl{}
\eqref{off-shell_quotient} for $\tau$-evolution, describes a dynamical
scalar field subjected to \eqref{nlin_KG}. The presence of the nonlinear
constraint, which cannot be manifestly resolved, complicates the analysis
of this system. But if one considers another, nonlinear, off-shell
scalar system, the constraint becomes linear and can be easily resolved,
leading directly to \eqref{F_eq} \cite{unfQFT}. This demonstrates
that, despite the formal equivalence of all off-shell formulations,
practical analysis requires the choice of a very specific one.

This way of generating a dynamical unfolded system from a non-dynamical
one is equivalent to the well-known method used for $d$-dimensional
unfolded theories, where unfolded fields are tensors. There, on-shell
factorization is implemented by imposing some trace constraints on
off-shell tensor fields \cite{ProkVas,unf3,ActionsCharges}. These
constraints may be very complicated: there are examples of nonlinear
off-shell HS theories \cite{ActionsCharges,did2}, whose explicit
on-shell reductions are still not available.

\section{Unfolded Scalar Electrodynamics\label{SEC_QED}}

In this Section, we construct an unfolded formulation for a self-interacting
complex scalar, minimally interacting with an electromagnetic field.
In the standard formulation, this corresponds to Lagrangian e.o.m.
\begin{equation}
\frac{1}{2}\mathrm{D}_{\alpha\dot{\alpha}}\mathrm{D}^{\alpha\dot{\alpha}}\phi+(m^{2}+\mathrm{U}'(\phi\phi^{*}))\phi=0,\quad\frac{1}{2}\mathrm{D}_{\alpha\dot{\alpha}}^{*}\mathrm{D}^{*\alpha\dot{\alpha}}\phi^{*}+(m^{2}+\mathrm{U}'(\phi\phi^{*}))\phi^{*}=0,\label{scalar_eom}
\end{equation}
\begin{equation}
\nabla_{\beta\dot{\alpha}}F^{\beta}{}_{\alpha}=iq(\phi\mathrm{D}_{\alpha\dot{\alpha}}^{*}\phi^{*}-\phi^{*}\mathrm{D}_{\alpha\dot{\alpha}}\phi),\quad\nabla_{\alpha\dot{\beta}}\bar{F}^{\dot{\beta}}{}_{\dot{\alpha}}=iq(\phi\mathrm{D}_{\alpha\dot{\alpha}}^{*}\phi^{*}-\phi^{*}\mathrm{D}_{\alpha\dot{\alpha}}\phi),\label{Maxwell_eom}
\end{equation}
where $q$ is an electric charge and covariant derivatives are defined
as
\begin{equation}
\mathrm{D}_{\alpha\dot{\alpha}}:=\nabla_{\alpha\dot{\alpha}}-iqA_{\alpha\dot{\alpha}},\quad\mathrm{D}_{\alpha\dot{\alpha}}^{*}:=\nabla_{\alpha\dot{\alpha}}+iqA_{\alpha\dot{\alpha}},\label{covar_derivative_def}
\end{equation}
\begin{equation}
[\mathrm{D}_{\alpha\dot{\alpha}},\mathrm{D}_{\beta\dot{\beta}}]=-iq\epsilon_{\alpha\beta}\bar{F}_{\dot{\alpha}\dot{\beta}}-iq\epsilon_{\dot{\alpha}\dot{\beta}}F_{\alpha\beta},\quad[\mathrm{D}_{\alpha\dot{\alpha}}^{*},\mathrm{D}_{\beta\dot{\beta}}^{*}]=iq\epsilon_{\alpha\beta}\bar{F}_{\dot{\alpha}\dot{\beta}}+iq\epsilon_{\dot{\alpha}\dot{\beta}}F_{\alpha\beta}.\label{covar_derivative_comm}
\end{equation}

Our strategy goes as follows. We start with unfolding a conserved
electric current of a general form. Then we couple this current to
the unfolded Maxwell field. Finally, we unfold equations for a charged
complex scalar field coupled to the Maxwell field and express the
electric current that we started with in terms of this unfolded scalar,
thus closing the system.

\subsection{Electric current}

We are to find an unfolded system describing a conserved electric
current of a general form, i.e. a vector field $j_{\alpha\dot{\alpha}}(x)$
subjected to the conservation condition
\begin{equation}
\nabla_{\alpha\dot{\alpha}}j^{\alpha\dot{\alpha}}=0.\label{current_conserv}
\end{equation}
We begin with postulating the form of the corresponding unfolded field
\begin{equation}
\mathrm{J}(Y|x):=e^{\nabla y\bar{y}}j_{\alpha\dot{\alpha}}y^{\alpha}\bar{y}^{\dot{\alpha}}.\label{J_postulate}
\end{equation}
Next, we need to calculate $\partial_{\alpha}\bar{\partial}_{\dot{\alpha}}\mathrm{J}$
and express it in terms of $\nabla_{\alpha\dot{\alpha}}\mathrm{J}$
and spinorial operators acting on unfolded fields. We have
\begin{equation}
\partial_{\alpha}\mathrm{J}=\nabla_{\alpha\dot{\beta}}\bar{y}^{\dot{\beta}}\mathrm{J}+e^{\nabla y\bar{y}}j_{\alpha\dot{\beta}}\bar{y}^{\dot{\beta}},
\end{equation}
\begin{equation}
\partial_{\alpha}\bar{\partial}_{\dot{\alpha}}\mathrm{J}=(N+2)\nabla_{\alpha\dot{\alpha}}\mathrm{J}-y_{\alpha}\bar{y}_{\dot{\alpha}}\square\mathrm{J}+e^{\nabla y\bar{y}}j_{\alpha\dot{\alpha}}+e^{\nabla y\bar{y}}\bar{y}_{\dot{\alpha}}\nabla_{\alpha\dot{\beta}}j_{\beta}{}^{\dot{\beta}}y^{\beta}+e^{\nabla y\bar{y}}y_{\alpha}\nabla_{\beta\dot{\alpha}}j^{\beta}{}_{\dot{\beta}}\bar{y}^{\dot{\beta}}.\label{ddJ_intermed}
\end{equation}
The third term on the r.h.s. can be rewritten as
\begin{equation}
e^{\nabla y\bar{y}}j_{\alpha\dot{\alpha}}=j_{\alpha\dot{\alpha}}+\frac{1}{N}(\nabla y\bar{y})e^{\nabla y\bar{y}}j_{\alpha\dot{\alpha}}=j_{\alpha\dot{\alpha}}+\frac{1}{N}\nabla_{\alpha\dot{\alpha}}\mathrm{J}+\frac{1}{2N}e^{\nabla y\bar{y}}(\bar{y}_{\dot{\alpha}}\partial_{\alpha}(\nabla_{\beta\dot{\beta}}j_{\beta}{}^{\dot{\beta}}y^{\beta}y^{\beta})+y_{\alpha}\bar{\partial}_{\dot{\alpha}}(\nabla_{\beta\dot{\beta}}j^{\beta}{}_{\dot{\beta}}\bar{y}^{\dot{\beta}}\bar{y}^{\dot{\beta}})),
\end{equation}
where we have taken into account that
\begin{equation}
\nabla_{\alpha\dot{\gamma}}j_{\beta}{}^{\dot{\gamma}}=\nabla_{\beta\dot{\gamma}}j_{\alpha}{}^{\dot{\gamma}}
\end{equation}
due to the conservation condition \eqref{current_conserv}. Next,
\begin{equation}
e^{\nabla y\bar{y}}\bar{y}_{\dot{\alpha}}\partial_{\alpha}(\nabla_{\beta\dot{\beta}}j_{\beta}{}^{\dot{\beta}}y^{\beta}y^{\beta})=\bar{y}_{\dot{\alpha}}\partial_{\alpha}\mathrm{J}^{+}-\bar{y}_{\dot{\alpha}}\nabla_{\alpha\dot{\gamma}}\bar{y}^{\dot{\gamma}}\mathrm{J}^{+},
\end{equation}
where a new unfolded field $\mathrm{J}^{+}$ is introduced as
\begin{equation}
\mathrm{J}^{+}(Y|x):=e^{\nabla y\bar{y}}\nabla_{\beta\dot{\beta}}j_{\beta}{}^{\dot{\beta}}y^{\beta}y^{\beta},\label{J+_def}
\end{equation}
with a complex conjugate unfolded field $\mathrm{J}^{-}$ being
\begin{equation}
\mathrm{J}^{-}(Y|x):=e^{\nabla y\bar{y}}\nabla_{\beta\dot{\beta}}j^{\beta}{}_{\dot{\beta}}\bar{y}^{\dot{\beta}}\bar{y}^{\dot{\beta}}.
\end{equation}
The reason for introducing additional unfolded fields is that there
are new sequences of differential descendants of $j_{\alpha\dot{\alpha}}$,
that neither fit into the sequence of symmetrized traceless derivatives
contained in $\mathrm{J}$ nor are fixed by the differential constraint
\eqref{current_conserv}. They arise from $\nabla_{\beta\dot{\beta}}j_{\beta}{}^{\dot{\beta}}$
and $\nabla_{\beta\dot{\beta}}j^{\beta}{}_{\dot{\beta}}$, which generate
$\mathrm{J}^{+}$ and $\mathrm{J}^{-}$ respectively.

For these new unfolded fields, we need unfolded equations as well.
We have
\begin{equation}
\bar{\partial}_{\dot{\alpha}}\mathrm{J}^{+}=\nabla_{\beta\dot{\alpha}}y^{\beta}\mathrm{J}^{+},
\end{equation}
\begin{equation}
\partial_{\alpha}\bar{\partial}_{\dot{\alpha}}\mathrm{J}^{+}=(N+1)\nabla_{\alpha\dot{\alpha}}\mathrm{J}^{+}-y_{\alpha}\bar{y}_{\dot{\alpha}}\square\mathrm{J}^{+}-2y_{\alpha}\bar{\partial}_{\dot{\alpha}}\square\mathrm{J}+2y_{\alpha}\nabla_{\beta\dot{\alpha}}y^{\beta}\square\mathrm{J}.\label{J+_intermed}
\end{equation}
From here we get, after contracting with $\bar{y}^{\dot{\alpha}}$,
\begin{equation}
\nabla_{\alpha\dot{\alpha}}\bar{y}^{\dot{\alpha}}\mathrm{J}^{+}=\frac{\bar{N}}{\bar{N}+2}\partial_{\alpha}\mathrm{J}^{+}+\frac{2}{\bar{N}+2}y_{\alpha}\square\mathrm{J}.
\end{equation}
This allows us to close the expression for $\partial_{\alpha}\bar{\partial}_{\dot{\alpha}}\mathrm{J}$
\eqref{ddJ_intermed}, except for the terms with $j_{\alpha\dot{\alpha}}$
and $\square\mathrm{J}$,
\begin{equation}
\partial_{\alpha}\bar{\partial}_{\dot{\alpha}}\mathrm{J}=j_{\alpha\dot{\alpha}}+\frac{(N+1)^{2}}{N}\nabla_{\alpha\dot{\alpha}}\mathrm{J}-\frac{(N+2)}{N}y_{\alpha}\bar{y}_{\dot{\alpha}}\square\mathrm{J}+\frac{1}{N}\bar{y}_{\dot{\alpha}}\partial_{\alpha}\mathrm{J}^{+}+\frac{1}{N}y_{\alpha}\bar{\partial}_{\dot{\alpha}}\mathrm{J}^{-}.\label{ddJ}
\end{equation}
Multiplying by $\frac{N}{(N+1)^{2}}$ and contracting with the vierbein,
an unfolded equation for $\mathrm{J}$ arises
\begin{equation}
\mathrm{d}_{L}\mathrm{J}-\frac{N}{(N+1)^{2}}e\partial\bar{\partial}\mathrm{J}-\frac{(N+2)}{(N+1)^{2}}ey\bar{y}\square\mathrm{J}+\frac{1}{(N+1)^{2}}e\bar{y}\partial\mathrm{J}^{+}+\frac{1}{(N+1)^{2}}ey\bar{\partial}\mathrm{J}^{-}=0,\label{J_eq}
\end{equation}
where
\begin{equation}
e\bar{y}\partial:=e^{\alpha\dot{\beta}}\bar{y}_{\dot{\beta}}\partial_{\alpha},\quad ey\bar{\partial}:=e^{\alpha\dot{\beta}}y_{\alpha}\bar{\partial}_{\dot{\beta}}.
\end{equation}
Note that $j_{\alpha\dot{\alpha}}$ drops out of \eqref{J_eq} because
$Nj_{\alpha\dot{\alpha}}=0$.

Now, to close the $\mathrm{J}^{+}$-equation \eqref{J+_intermed},
we contract \eqref{ddJ} with $y^{\alpha}\frac{N}{(N+1)^{2}}$, that
produces
\begin{equation}
\nabla_{\beta\dot{\alpha}}y^{\beta}\mathrm{J}=\frac{(N-1)}{N}\bar{\partial}_{\dot{\alpha}}\mathrm{J}-\frac{1}{N}\bar{y}_{\dot{\alpha}}\mathrm{J}^{+},\label{J_J+}
\end{equation}
so that
\begin{equation}
\partial_{\alpha}\bar{\partial}_{\dot{\alpha}}\mathrm{J}^{+}=(N+1)\nabla_{\alpha\dot{\alpha}}\mathrm{J}^{+}-\frac{(N+1)}{(N-1)}y_{\alpha}\bar{y}_{\dot{\alpha}}\square\mathrm{J}^{+}-2\frac{1}{(N-1)}y_{\alpha}\bar{\partial}_{\dot{\alpha}}\square\mathrm{J},
\end{equation}
and the unfolded equation for $\mathrm{J}^{+}$ is
\begin{equation}
\mathrm{d}_{L}\mathrm{J}^{+}-\frac{1}{(N+1)}e\partial\bar{\partial}\mathrm{J}^{+}-\frac{1}{(\bar{N}+1)}ey\bar{y}\square\mathrm{J}^{+}-\frac{2}{(N+1)(\bar{N}+1)}ey\bar{\partial}\square\mathrm{J}=0.\label{J+_eq}
\end{equation}
The equation \eqref{J_J+} also leads to a simple relation between
$\mathrm{J}$ and $\mathrm{J}^{+}$
\begin{equation}
\mathrm{J}^{+}=y^{\alpha}\bar{\partial}^{\dot{\alpha}}\nabla_{\alpha\dot{\alpha}}\mathrm{J}.\label{J+_dJ}
\end{equation}
Conjugation gives an unfolded equation for $\mathrm{J}^{-}$
\begin{equation}
\mathrm{d}_{L}\mathrm{J}^{-}-\frac{1}{(\bar{N}+1)}e\partial\bar{\partial}\mathrm{J}^{-}-\frac{1}{(N+1)}ey\bar{y}\square\mathrm{J}^{-}-\frac{2}{(N+1)(\bar{N}+1)}e\bar{y}\partial\square\mathrm{J}=0\label{J-_eq}
\end{equation}
and
\begin{equation}
\mathrm{J}^{-}=\bar{y}^{\dot{\alpha}}\partial^{\alpha}\nabla_{\alpha\dot{\alpha}}\mathrm{J}.\label{J-_dJ}
\end{equation}

This solves the problem: the system of three unfolded equations \eqref{J_eq},
\eqref{J+_eq} and \eqref{J-_eq} represents an unfolded formulation
for the electric current, with the solution being \eqref{J_postulate},
\eqref{J+_dJ}, \eqref{J-_dJ} and with the primary field obeying
\eqref{current_conserv}. Strictly speaking, in order to have a completely
unfolded system, one also needs to process $\square$-terms, because
the spacetime derivative is allowed to appear in an unfolded equation
only through the exterior derivative. This can be done either by introducing
one more auxiliary (scalar) variable on top of spinors $Y$ that encodes
new unfolded descendants \cite{unf8}, if the electric current is
off-shell, or by expressing them in terms of $\mathrm{J}$, if the
current is built out of some dynamical on-shell fields. But here we
leave the equations as they are for now, since this system plays only
an intermediate role in our analysis.

The system can be made more compact via defining a united unfolded
master-field as
\begin{equation}
\mathcal{J}(Y|x):=\mathrm{J}+\mathrm{J}^{+}+\mathrm{J}^{-}.
\end{equation}
Then, introducing an averaged Euler operator
\begin{equation}
\nu:=\frac{N+\bar{N}}{2},
\end{equation}
three equations \eqref{J_eq}, \eqref{J+_eq}, \eqref{J-_eq} can
be combined into one
\begin{equation}
\mathrm{d}_{L}\mathcal{J}-\frac{1}{(N+1)(\bar{N}+1)}\left\{ \nu e\partial\bar{\partial}\mathcal{J}+(\nu+2)ey\bar{y}\square\mathcal{J}-e\bar{y}\partial(\mathrm{J}^{+}-2\square\mathcal{J})-ey\bar{\partial}(\mathrm{J}^{-}-2\square\mathcal{J})\right\} =0.\label{J_united_eq}
\end{equation}

\subsection{Maxwell equations}

The second step is to construct an unfolded system for the electromagnetic
field sourced by the unfolded electric current.

We need to unfold Maxwell equations
\begin{equation}
\nabla_{\beta\dot{\alpha}}F^{\beta}{}_{\alpha}=qj_{\alpha\dot{\alpha}},\quad\nabla_{\alpha\dot{\beta}}\bar{F}^{\dot{\beta}}{}_{\dot{\alpha}}=qj_{\alpha\dot{\alpha}},\label{Maxwell_eq}
\end{equation}
where the (anti-)selfdual Maxwell tensor is determined by the gauge
field $A_{\alpha\dot{\alpha}}(x)$ as
\begin{equation}
F_{\alpha\alpha}:=\nabla_{\alpha\dot{\beta}}A_{\alpha}{}^{\dot{\beta}},\quad\bar{F}_{\dot{\alpha}\dot{\alpha}}:=\nabla_{\beta\dot{\alpha}}A^{\beta}{}_{\dot{\alpha}}.\label{Maxwell_tensor_def}
\end{equation}

To make a gauge symmetry manifest, we treat $A_{\alpha\dot{\alpha}}(x)$
as the vierbein expansion of an unfolded 1-form $A$ 
\begin{equation}
A(x)=e^{\alpha\dot{\alpha}}A_{\alpha\dot{\alpha}},
\end{equation}
that will generate unfolded $U(1)$ gauge symmetry in the final system
according to the general formula \eqref{unf_gauge_transf}.

Now we postulate following expressions for the unfolded 0-forms of
the Maxwell tensors
\begin{equation}
F(Y|x)=e^{\nabla y\bar{y}}F_{\alpha\alpha}y^{\alpha}y^{\alpha},\quad\bar{F}(Y|x)=e^{\nabla y\bar{y}}\bar{F}_{\dot{\alpha}\dot{\alpha}}\bar{y}^{\dot{\alpha}}\bar{y}^{\dot{\alpha}}.\label{Maxwell_postulate}
\end{equation}
Then \eqref{Maxwell_tensor_def} can be written in the unfolded form
as
\begin{equation}
\mathrm{d}A=\frac{1}{4}e^{\alpha}{}_{\dot{\beta}}e^{\alpha\dot{\beta}}\partial_{\alpha}\partial_{\alpha}F|_{\bar{y}=0}+\frac{1}{4}e_{\beta}{}^{\dot{\alpha}}e^{\beta\dot{\alpha}}\bar{\partial}_{\dot{\alpha}}\bar{\partial}_{\dot{\alpha}}\bar{F}|_{y=0}.\label{dA_eef}
\end{equation}

The next task is to calculate and process $\partial_{\alpha}\bar{\partial}_{\dot{\alpha}}F$.
We get
\begin{equation}
\bar{\partial}_{\dot{\alpha}}F=\nabla_{\beta\dot{\alpha}}y^{\beta}F,
\end{equation}
\begin{equation}
\partial_{\alpha}\bar{\partial}_{\dot{\alpha}}F=(N+1)\nabla_{\alpha\dot{\alpha}}F-y_{\alpha}\bar{y}_{\dot{\alpha}}\square F+2qy_{\alpha}e^{\nabla y\bar{y}}j_{\beta\dot{\alpha}}y^{\beta},\label{ddF_intermediate}
\end{equation}
where we made use of Maxwell equations \eqref{Maxwell_eq}. They also
yield
\begin{equation}
\square F_{\alpha\beta}=-q\nabla_{\alpha\dot{\beta}}j_{\beta}{}^{\dot{\beta}},
\end{equation}
so that, taking into account \eqref{J+_def},
\begin{equation}
\square F=-q\mathrm{J}^{+}.
\end{equation}
The last term in \eqref{ddF_intermediate} can be re-expressed as
\begin{equation}
y_{\alpha}e^{\nabla y\bar{y}}j_{\beta\dot{\alpha}}y^{\beta}=y_{\alpha}\bar{\partial}_{\dot{\alpha}}\mathrm{J}-y_{\alpha}y^{\beta}\nabla_{\beta\dot{\alpha}}\mathrm{J}.
\end{equation}
The equation \eqref{ddJ} leads to
\begin{equation}
y^{\beta}\nabla_{\beta\dot{\alpha}}\mathrm{J}=\frac{(N-1)}{(\bar{N}+1)}\bar{\partial}_{\dot{\alpha}}\mathrm{J}-\frac{1}{(\bar{N}+1)}\bar{y}_{\dot{\alpha}}\mathrm{J}^{+},
\end{equation}
hence
\begin{equation}
\partial_{\alpha}\bar{\partial}_{\dot{\alpha}}F=(N+1)\nabla_{\alpha\dot{\alpha}}F+q\frac{(N+1)}{(\bar{N}+1)}y_{\alpha}\bar{y}_{\dot{\alpha}}\mathrm{J}^{+}+\frac{2q}{(\bar{N}+1)}y_{\alpha}\bar{\partial}_{\dot{\alpha}}\mathrm{J},\label{ddF_final}
\end{equation}
which solves the problem. Contracting this with the vierbein, we obtain
the required unfolded equation
\begin{equation}
\mathrm{d}_{L}F-\frac{1}{(N+1)(\bar{N}+1)}\left\{ \nu e\partial\bar{\partial}F-q(\nu+2)ey\bar{y}\mathrm{J}^{+}-2qey\bar{\partial}\mathrm{J}\right\} =0.\label{Maxwell_eq_gen_J}
\end{equation}
Conjugation gives
\begin{equation}
\mathrm{d}_{L}\bar{F}-\frac{1}{(N+1)(\bar{N}+1)}\left\{ \nu e\partial\bar{\partial}\bar{F}-q(\nu+2)ey\bar{y}\mathrm{J}^{-}-2qe\bar{y}\partial\mathrm{J}\right\} =0.\label{Maxwell_eq_gen_J_*}
\end{equation}
From \eqref{ddF_final} also a simple expression for $\mathrm{J}$
follows
\begin{equation}
\mathrm{J}=-\frac{1}{2q}\bar{y}^{\dot{\alpha}}\partial^{\alpha}\nabla_{\alpha\dot{\alpha}}F.
\end{equation}
This implies that, from a formally-mathematical point of view, the
unfolded electric current $\mathrm{J}$ represents just a subsequence
of unfolded descendants of $A$. Indeed, Maxwell equations \eqref{Maxwell_eq}
can be interpreted as a constraint, which expresses the descendant
field $j_{\alpha\dot{\alpha}}$ in terms of the primary field $F_{\alpha\alpha}$.
This simple observation plays an important role in the quantization
of unfolded field theories \cite{unfQFT}.

\subsection{Charged scalar field}

Now we move to the main part of the problem: constructing an unfolded
system for the charged scalar field, interacting with the electromagnetic
one. As was said above, while the cases of electric current and Maxwell
field, representing linear models, can be solved via direct consistency
consideration \cite{unf8}, this looks almost unfeasible for the problem
at hand.

We need to unfold equations \eqref{scalar_eom}. We postulate unfolded
scalar fields to be
\begin{equation}
\Phi(Y|x)=e^{y^{\beta}\bar{y}^{\dot{\beta}}\mathrm{D_{\beta\dot{\beta}}}}\phi(x),\quad\Phi^{*}(Y|x)=e^{y^{\beta}\bar{y}^{\dot{\beta}}\mathrm{D_{\beta\dot{\beta}}^{*}}}\phi^{*}(x).\label{Phi_postulate}
\end{equation}
This unfolding is strongly nonlinear, containing the exponent of the
dynamical gauge field $A_{\alpha\dot{\alpha}}$, as opposite to all
unfoldings considered before. The reason behind this is that we want
to keep the gauge invariance manifest. This means that the unfolded
fields should transform covariantly, which forces one to replace ordinary
derivatives in the exponent, present in the non-gauge case \eqref{F_postulate},
with the covariant ones. Then $\partial_{\alpha}\bar{\partial}_{\dot{\alpha}}\Phi$
generates a covariant term $\mathrm{D_{\alpha\dot{\alpha}}}\Phi$,
which gives rise to $U(1)$ gauge symmetry according to the general
formula \eqref{unf_gauge_transf}.

From \eqref{Phi_postulate} a useful relation immediately follows
\begin{equation}
y^{\beta}\bar{y}^{\dot{\beta}}\mathrm{D_{\beta\dot{\beta}}}\Phi=N\Phi=\bar{N}\Phi.\label{yyDPhi_NPhi}
\end{equation}
We simplify the appearance of formulas by writing
\begin{equation}
\mathrm{D}y\bar{y}:=y^{\beta}\bar{y}^{\dot{\beta}}\mathrm{D_{\beta\dot{\beta}}},\quad\mathrm{D}^{*}y\bar{y}:=y^{\beta}\bar{y}^{\dot{\beta}}\mathrm{D_{\beta\dot{\beta}}^{*}},\quad\mathrm{D}^{2}:=\frac{1}{2}\mathrm{D}_{\beta\dot{\beta}}\mathrm{D}^{\beta\dot{\beta}},\quad\mathrm{D}^{*2}:=\frac{1}{2}\mathrm{D}_{\beta\dot{\beta}}^{*}\mathrm{D}^{*\beta\dot{\beta}}.
\end{equation}

Now we start calculating $\partial_{\alpha}\bar{\partial}_{\dot{\alpha}}\Phi$.
Combining a relation for a commutator
\begin{equation}
[\hat{\mathrm{A}},e^{\hat{\mathrm{D}}}]=\int_{0}^{1}dte^{t\hat{\mathrm{D}}}[\hat{\mathrm{A}},\hat{\mathrm{D}}]e^{-t\hat{\mathrm{D}}}e^{\hat{\mathrm{D}}}\label{commutator}
\end{equation}
with an Euler-operator representation of a homotopy integral
\begin{equation}
\int_{0}^{1}dtt^{k}F(tz)=\frac{1}{z\frac{\partial}{\partial z}+1+k}F(z),\label{homotopy}
\end{equation}
we get
\begin{equation}
\partial_{\alpha}\Phi=\mathrm{D_{\alpha\dot{\beta}}}\bar{y}^{\dot{\beta}}\Phi-y_{\alpha}\Phi\frac{iq}{(N+1)(N+2)}\bar{F},\label{partial_d_F}
\end{equation}
\begin{eqnarray}
 &  & \partial_{\alpha}\bar{\partial}_{\dot{\alpha}}\Phi=(N+1)\mathrm{D_{\alpha\dot{\alpha}}}\Phi-y_{\alpha}\bar{\partial}_{\dot{\alpha}}(\Phi\frac{iq}{(N+1)(N+2)}\bar{F})-\frac{iq}{2}\Phi\bar{y}_{\dot{\alpha}}\partial_{\alpha}F_{\beta\beta}y^{\beta}y^{\beta}-y_{\alpha}\bar{y}_{\dot{\alpha}}\mathrm{D}^{2}F-\nonumber \\
 &  & -\bar{y}_{\dot{\alpha}}\Phi\frac{iq}{\bar{N}(\bar{N}+1)}\bar{y}^{\dot{\beta}}\nabla_{\alpha\dot{\beta}}F-\bar{y}_{\dot{\alpha}}(\frac{iq}{(\bar{N}+1)(\bar{N}+2)}F)(\partial_{\alpha}\Phi+y_{\alpha}\Phi\frac{iq}{(N+1)(N+2)}\bar{F}).
\end{eqnarray}
From \eqref{ddF_final} one finds
\begin{equation}
\nabla_{\alpha\dot{\beta}}\bar{y}^{\dot{\beta}}F=\partial_{\alpha}\frac{\bar{N}}{N}F-\frac{2q}{(N+1)}y_{\alpha}\mathrm{J},
\end{equation}
so that
\begin{eqnarray}
 &  & \partial_{\alpha}\bar{\partial}_{\dot{\alpha}}\Phi=(N+1)\mathrm{D_{\alpha\dot{\alpha}}}\Phi-y_{\alpha}\bar{\partial}_{\dot{\alpha}}(\Phi\frac{iq}{(N+1)(N+2)}\bar{F})-\bar{y}_{\dot{\alpha}}\partial_{\alpha}(\Phi\frac{iq}{(\bar{N}+1)(\bar{N}+2)}F)+\nonumber \\
 &  & +q^{2}y_{\alpha}\bar{y}_{\dot{\alpha}}\Phi(\frac{1}{(\bar{N}+1)(\bar{N}+2)}F)(\frac{1}{(N+1)(N+2)}\bar{F})+y_{\alpha}\bar{y}_{\dot{\alpha}}\Phi\frac{2iq^{2}}{N(N+1)(N+2)}\mathrm{J}-y_{\alpha}\bar{y}_{\dot{\alpha}}\mathrm{D}^{2}\Phi.\nonumber \\
\label{ddPhi_initial}
\end{eqnarray}
This might seem like a solution to the problem. However, the last
term is not of acceptable form. The situation here is different from
the non-gauge case \eqref{ddPhi_non_gauge}, since the covariant box
$\mathrm{D}^{2}$ does not commute with the unfolding exponent $\exp(\mathrm{D}y\bar{y})$
and one cannot directly apply e.o.m. for the primary field \eqref{scalar_eom}.

By means of the same combination of the commutator \eqref{commutator}
and the homotopy integral \eqref{homotopy}, we can write an unfolded
generalization for \eqref{scalar_eom} as
\begin{equation}
\mathrm{D}^{2}\Phi=-(m^{2}+\mathrm{U}'(\Phi\Phi^{*}))\Phi+(\frac{1}{N}e^{\mathrm{D}y\bar{y}}[\mathrm{D}^{2},\mathrm{D}y\bar{y}]e^{-\mathrm{D}y\bar{y}})\Phi.\label{DDPhi_start}
\end{equation}
The commutator in \eqref{DDPhi_start} can be expanded as
\begin{equation}
[\mathrm{D}^{2},\mathrm{D}y\bar{y}]=\frac{1}{2}[\mathrm{D}_{\beta\dot{\beta}},\mathrm{D}y\bar{y}]\mathrm{D}^{\beta\dot{\beta}}+\frac{1}{2}\mathrm{D}^{\beta\dot{\beta}}[\mathrm{D}_{\beta\dot{\beta}},\mathrm{D}y\bar{y}]=iq(y_{\beta}\bar{F}_{\dot{\alpha}\dot{\beta}}\bar{y}^{\dot{\alpha}}+\bar{y}_{\dot{\beta}}F_{\alpha\beta}y^{\alpha})\mathrm{D}^{\beta\dot{\beta}}+iq^{2}j_{\alpha\dot{\alpha}}y^{\alpha}\bar{y}^{\dot{\alpha}},
\end{equation}
where \eqref{covar_derivative_comm} and \eqref{Maxwell_eq} are used,
so that
\begin{equation}
\frac{1}{N}e^{\mathrm{D}y\bar{y}}[\mathrm{D}^{2},\mathrm{D}y\bar{y}]e^{-\mathrm{D}y\bar{y}}=iq^{2}\frac{1}{N}\mathrm{J}+iq\frac{1}{N}e^{\mathrm{D}y\bar{y}}(\bar{y}_{\dot{\beta}}F_{\alpha\beta}y^{\alpha}+y_{\beta}\bar{F}_{\dot{\alpha}\dot{\beta}}\bar{y}^{\dot{\alpha}})\mathrm{D}^{\beta\dot{\beta}}e^{-\mathrm{D}y\bar{y}}.
\end{equation}
Next,
\begin{equation}
e^{\mathrm{D}y\bar{y}}\bar{y}_{\dot{\beta}}F_{\alpha\beta}y^{\alpha}\mathrm{D}^{\beta\dot{\beta}}e^{-\mathrm{D}y\bar{y}}=(e^{\nabla y\bar{y}}(\frac{1}{2}\partial_{\beta}F_{\alpha\alpha}y^{\alpha}y^{\alpha})e^{-\nabla y\bar{y}})(e^{\mathrm{D}y\bar{y}}\bar{y}_{\dot{\beta}}\mathrm{D}^{\beta\dot{\beta}}e^{-\mathrm{D}y\bar{y}}),
\end{equation}
and two factors can be rewritten as
\begin{equation}
e^{\nabla y\bar{y}}(\partial_{\beta}F_{\alpha\alpha}y^{\alpha}y^{\alpha})e^{-\nabla y\bar{y}}=\frac{2}{(N+1)}\partial_{\beta}F+\frac{2q}{(N+1)}y_{\beta}\mathrm{J},
\end{equation}
\begin{equation}
e^{\mathrm{D}y\bar{y}}\bar{y}_{\dot{\beta}}\mathrm{D}^{\beta\dot{\beta}}e^{-\mathrm{D}y\bar{y}}=\bar{y}_{\dot{\beta}}\mathrm{D}^{\beta\dot{\beta}}+iq\frac{1}{N}y^{\beta}\bar{F}.
\end{equation}
This way we get
\begin{eqnarray}
 &  & \mathrm{D}^{2}\Phi=-(m^{2}+\mathrm{U}'(\Phi\Phi^{*}))\Phi+iq^{2}\Phi\frac{1}{N}\mathrm{J}+(\frac{2iq^{2}}{(N+1)(N+2)}\mathrm{J})y^{\alpha}\bar{y}^{\dot{\alpha}}\mathrm{D}_{\alpha\dot{\alpha}}\Phi-\nonumber \\
 &  & -\Phi\frac{iq}{N}(F\frac{1}{N+1}\bar{F}+\bar{F}\frac{1}{\bar{N}+1}F)+(\frac{iq}{N(N+1)}\partial_{\alpha}F)\bar{y}_{\dot{\alpha}}\mathrm{D}^{\alpha\dot{\alpha}}\Phi+(\frac{iq}{\bar{N}(\bar{N}+1)}\bar{\partial}_{\dot{\alpha}}\bar{F})y_{\alpha}\mathrm{D}^{\alpha\dot{\alpha}}\Phi,\nonumber \\
\end{eqnarray}
and, making use of \eqref{yyDPhi_NPhi} and \eqref{partial_d_F} (and
its conjugate), arrive at the required representation
\begin{eqnarray}
 &  & \mathrm{D}^{2}\Phi=-(m^{2}+\mathrm{U}'(\Phi\Phi^{*}))\Phi+\Phi\frac{iq^{2}}{N}\mathrm{J}+(\frac{2iq^{2}}{(N+1)(N+2)}\mathrm{J})N\Phi-\Phi\frac{iq}{N}(F\frac{1}{N+1}\bar{F}+\bar{F}\frac{1}{\bar{N}+1}F)-\nonumber \\
 &  & -(\frac{iq}{N(N+1)}\partial_{\alpha}F)(\partial^{\alpha}\Phi+y^{\alpha}\Phi\frac{iq}{(N+1)(N+2)}\bar{F})-(\frac{iq}{\bar{N}(\bar{N}+1)}\bar{\partial}_{\dot{\alpha}}\bar{F})(\bar{\partial}^{\dot{\alpha}}\Phi+\bar{y}^{\dot{\alpha}}\Phi\frac{iq}{(\bar{N}+1)(\bar{N}+2)}F).\nonumber \\
\label{D2_Phi}
\end{eqnarray}
Substituting this into \eqref{ddPhi_initial} yields $\partial_{\alpha}\bar{\partial}_{\dot{\alpha}}\Phi$
in a suitable form
\begin{eqnarray}
 &  & \partial_{\alpha}\bar{\partial}_{\dot{\alpha}}\Phi=(N+1)\mathrm{D_{\alpha\dot{\alpha}}}\Phi+y_{\alpha}\bar{y}_{\dot{\alpha}}(m^{2}+\mathrm{U}'(\Phi\Phi^{*}))\Phi-\nonumber \\
 &  & -iqy_{\alpha}\bar{\partial}_{\dot{\alpha}}(\Phi\frac{1}{(N+1)(N+2)}\bar{F})-iq(y_{\alpha}\bar{\partial}_{\dot{\alpha}}\frac{1}{(N+1)(N+2)}\bar{F})N\Phi+iq(\frac{1}{(N+1)}\bar{F})y_{\alpha}\bar{\partial}_{\dot{\alpha}}\Phi-\nonumber \\
 &  & -iq\bar{y}_{\dot{\alpha}}\partial_{\alpha}(\Phi\frac{1}{(\bar{N}+1)(\bar{N}+2)}F)-iq(\bar{y}_{\dot{\alpha}}\partial_{\alpha}\frac{1}{(\bar{N}+1)(\bar{N}+2)}F)N\Phi+iq(\frac{1}{(\bar{N}+1)}F)\bar{y}_{\dot{\alpha}}\partial_{\alpha}\Phi+\nonumber \\
 &  & +q^{2}y_{\alpha}\bar{y}_{\dot{\alpha}}\Phi(\frac{1}{(\bar{N}+1)(\bar{N}+2)}F)(\frac{1}{(N+1)(N+2)}\bar{F})+q^{2}y_{\alpha}\bar{y}_{\dot{\alpha}}\Phi\frac{1}{N}(F\frac{1}{(N+1)}\bar{F}+\bar{F}\frac{1}{(\bar{N}+1)}F)-\nonumber \\
 &  & -q^{2}y_{\alpha}\bar{y}_{\dot{\alpha}}\Phi(\frac{1}{(\bar{N}+1)}F\cdot\frac{1}{(N+1)(N+2)}\bar{F}+\frac{1}{(N+1)}\bar{F}\cdot\frac{1}{(\bar{N}+1)(\bar{N}+2)}F)-\nonumber \\
 &  & -iq^{2}y_{\alpha}\bar{y}_{\dot{\alpha}}(-2\Phi\frac{1}{N(N+1)(N+2)}\mathrm{J}+\Phi\frac{1}{N}\mathrm{J}+2N\Phi\cdot\frac{1}{(N+1)(N+2)}\mathrm{J}).\label{ddPhi_final}
\end{eqnarray}

\subsection{Unfolded equations of scalar electrodynamics}

In \eqref{ddPhi_final} and in \eqref{Maxwell_eq_gen_J}-\eqref{Maxwell_eq_gen_J_*}
$\mathrm{J}$ is an arbitrary (e.g. some external) conserved electric
current for now. But in scalar electrodynamics \eqref{scalar_eom}-\eqref{Maxwell_eom}
it is made up of the dynamical scalar field and describes its back-reaction,
\begin{equation}
j_{\alpha\dot{\alpha}}=i(\phi^{*}\mathrm{D}_{\alpha\dot{\alpha}}\phi-\phi\mathrm{D}_{\alpha\dot{\alpha}}^{*}\phi^{*}),\quad\nabla^{\alpha\dot{\alpha}}j_{\alpha\dot{\alpha}}=0.\label{j_scalar}
\end{equation}
Applying unfolding \eqref{Phi_postulate} to this expression and comparing
with the unfolding of the current \eqref{J_postulate}, one finds
a surprisingly simple expression for an unfolded electric current
of the scalar field
\begin{equation}
\mathrm{J}=i(\Phi^{*}N\Phi-\Phi N\Phi^{*}).\label{J_scalar}
\end{equation}
Then from \eqref{J+_dJ}, \eqref{J-_dJ} it follows that
\begin{equation}
\mathrm{J}^{+}=-i\bar{\partial}_{\dot{\alpha}}\Phi\cdot\bar{\partial}^{\dot{\alpha}}\Phi^{*}-(\bar{N}+2)(\Phi\Phi^{*}\frac{2q}{(\bar{N}+2)}F),\quad\mathrm{J}^{-}=-i\partial_{\alpha}\Phi\cdot\partial^{\alpha}\Phi^{*}-(N+2)(\Phi\Phi^{*}\frac{2q}{(N+2)}\bar{F}).\label{J+_J-_scalar}
\end{equation}
Let us stress that this unfolded electric current is conserved on
nonlinear equations for the constituent scalar fields, which distinguishes
it from unfolded spin-1 conserved currents considered (together with
HS currents) e.g. in \cite{current1,current2,current3,current4},
where they are built up from free constituent fields.

Now, substituting \eqref{J_scalar}-\eqref{J+_J-_scalar} into \eqref{Maxwell_eq_gen_J}
and \eqref{Maxwell_eq_gen_J_*}, contracting \eqref{ddPhi_final}
with the vierbein and introducing 1-forms of covariant derivatives
as
\begin{equation}
\mathrm{D}:=e^{\alpha\dot{\alpha}}\mathrm{D}_{\alpha\dot{\alpha}}=\mathrm{d}_{L}-iqA,\quad\mathrm{D}^{*}:=e^{\alpha\dot{\alpha}}\mathrm{D}_{\alpha\dot{\alpha}}^{*}=\mathrm{d}_{L}+iqA,
\end{equation}
one finds after some algebraic simplifications in \eqref{ddPhi_final}

\begin{eqnarray}
 &  & \mathrm{d}A=\frac{1}{4}e^{\alpha}{}_{\dot{\beta}}e^{\alpha\dot{\beta}}\partial_{\alpha}\partial_{\alpha}F|_{\bar{y}=0}+\frac{1}{4}e_{\beta}{}^{\dot{\alpha}}e^{\beta\dot{\alpha}}\bar{\partial}_{\dot{\alpha}}\bar{\partial}_{\dot{\alpha}}\bar{F}|_{y=0},\label{dA_QED}\\
 &  & \mathrm{d}_{L}F-\frac{1}{(N+1)(\bar{N}+1)}\biggl\{\nu e\partial\bar{\partial}F-q(\nu+2)ey\bar{y}(i\bar{\partial}_{\dot{\alpha}}\Phi\cdot\bar{\partial}^{\dot{\alpha}}\Phi^{*}+(\bar{N}+2)(\Phi\Phi^{*}\frac{2q}{(\bar{N}+2)}F))-\nonumber \\
 &  & -2iqey\bar{\partial}(\Phi^{*}N\Phi-\Phi N\Phi^{*})\biggr\}=0,\label{Maxwell_eq_QED}
\end{eqnarray}
\begin{eqnarray}
 &  & (N+1)\mathrm{D}\Phi-e\partial\bar{\partial}\Phi+ey\bar{y}(m^{2}+\mathrm{U}'(\Phi\Phi^{*}))\Phi+q^{2}ey\bar{y}\Phi\cdot\frac{1}{(\bar{N}+2)}F\cdot\frac{1}{(N+2)}\bar{F}+\nonumber \\
 &  & +q^{2}ey\bar{y}\left(\Phi\frac{1}{(N+2)}(\Phi^{*}N\Phi-\Phi N\Phi^{*})+(N+1)\Phi\cdot\frac{2}{(N+1)(N+2)}(\Phi^{*}N\Phi-\Phi N\Phi^{*})\right)+\nonumber \\
 &  & +iq\left(\frac{1}{(\bar{N}+2)}F\cdot e\bar{y}\partial\Phi-(\bar{N}+1)\Phi\cdot e\bar{y}\partial\frac{1}{(\bar{N}+1)(\bar{N}+2)}F\right)+\nonumber \\
 &  & +iq\left(\frac{1}{(N+2)}\bar{F}\cdot ey\bar{\partial}\Phi-(N+1)\Phi\cdot ey\bar{\partial}\frac{1}{(N+1)(N+2)}\bar{F}\right)=0,\label{Phi_eq_QED}
\end{eqnarray}
plus an equation for $\bar{F}$ resulting from exchanging barred and
unbarred objects in \eqref{Maxwell_eq_QED} and an equation for $\Phi^{*}$
resulting from exchanging $\Phi$ and $\Phi^{*}$ and changing the
sign of $q$ in \eqref{Phi_eq_QED}. These equations form an unfolded
formulation of scalar electrodynamics. By construction, their solutions
(at least locally) are \eqref{Maxwell_tensor_def}, \eqref{Maxwell_postulate},
\eqref{Phi_postulate} with primary fields subjected to \eqref{scalar_eom}-\eqref{Maxwell_eom}.

It is interesting that in \eqref{Phi_eq_QED} there are cubic terms
in the second line that consist of the scalar field only. Traced back
to \eqref{D2_Phi}, these terms might seem like charged-current interactions.
However, they are not real vertices, as these terms do not contribute
to the e.o.m. of primary fields \eqref{scalar_eom}-\eqref{Maxwell_eom}.
Instead, their appearance must be attributed to the non-linearity
of the unfolding \eqref{Phi_postulate}, because it is differentiation
of the unfolding exponent, containing covariant derivatives, that
gives rise to them. The only scalar self-interaction is stored in
the potential $\mathrm{U}(\Phi\Phi^{*})$.

Making use of the general formula \eqref{unf_gauge_transf}, one obtains
manifest $U(1)$ gauge symmetry of the unfolded system \eqref{dA_QED}-\eqref{Phi_eq_QED}
with a gauge parameter $\varepsilon(x)$
\begin{equation}
\delta A=\mathrm{d}\varepsilon,\label{A_transform}
\end{equation}
\begin{equation}
\delta F=0,\quad\delta\bar{F}=0,\label{F_transform}
\end{equation}
\begin{equation}
\delta\Phi=iq\varepsilon\Phi,\quad\delta\Phi^{*}=-iq\varepsilon\Phi^{*}.\label{Phi_transform}
\end{equation}

\section{Higgs Mechanism\label{SEC_Higgs}}

Having the system \eqref{dA_QED}-\eqref{Phi_eq_QED} in hand, we
can study a realization of spontaneous $U(1)$ symmetry breaking in
the unfolded dynamics approach.

In the standard Lagrangian approach, one usually considers the massless
case with the \textquotedbl Mexican hat\textquotedbl{} scalar potential
\begin{equation}
\mathrm{U}(\phi\phi^{*})=-\mu^{2}\phi\phi^{*}+\frac{\lambda}{2}(\phi\phi^{*})^{2},\quad m^{2}=0.
\end{equation}
Then a continuum of classical vacuums is determined by
\begin{equation}
|\phi_{0}|=(\frac{\mu^{2}}{\lambda})^{1/2}.
\end{equation}
Considering some particular solution, e.g.
\begin{equation}
\phi_{0}=\frac{\mu}{\sqrt{\lambda}},\label{higgs_vacuum}
\end{equation}
and analyzing linear fluctuations over it in terms of two real scalar
fields
\begin{equation}
\phi(x)=\phi_{0}+\chi(x)+i\theta(x),
\end{equation}
one finds that the spectrum consists of one real massive scalar field
(Higgs boson) and a massive vector field, with the imaginary scalar
component $\theta$ eaten by the photon in order to gain weight.

Let us perform this analysis in terms of the unfolded system \eqref{dA_QED}-\eqref{Phi_eq_QED}.
We consider linearization in terms of
\begin{equation}
\Phi=\phi_{0}+\mathrm{X}+i\Theta,
\end{equation}
with the vacuum \eqref{higgs_vacuum}, so that
\begin{equation}
\mathrm{\nabla_{\alpha\dot{\alpha}}}\phi_{0}=N\phi_{0}=\bar{N}\phi_{0}=0.
\end{equation}
Substituting this into the unfolded equations \eqref{Maxwell_eq_QED}-\eqref{Phi_eq_QED}
gives in the linear limit
\begin{eqnarray}
 &  & \mathrm{d}_{L}F-\frac{1}{(N+1)(\bar{N}+1)}\left\{ \nu e\partial\bar{\partial}F-(\nu+2)2q^{2}\phi_{0}^{2}ey\bar{y}F+4q\phi_{0}ey\bar{\partial}N\Theta\right\} =0,\label{higgs_F_eq}\\
 &  & (N+1)\mathrm{d}_{L}\mathrm{X}-e\partial\bar{\partial}\mathrm{X}+2\mu^{2}ey\bar{y}\mathrm{X}=0,\label{higgs_X_eq}\\
 &  & (N+1)\mathrm{d}_{L}\Theta-q\phi_{0}A-e\partial\bar{\partial}\Theta+2q^{2}\phi_{0}^{2}ey\bar{y}\frac{N(N+3)}{(N+1)(N+2)}\Theta-\nonumber \\
 &  & -q\phi_{0}e\bar{y}\partial\frac{1}{(\bar{N}+1)(\bar{N}+2)}F-q\phi_{0}ey\bar{\partial}\frac{1}{(N+1)(N+2)}\bar{F}=0.\label{higgs_Theta_eq}
\end{eqnarray}
Equation \eqref{higgs_X_eq} indeed describes the Higgs boson with
the correct mass value
\begin{equation}
m_{X}^{2}=2\mu^{2}.
\end{equation}
To see that the rest of equations describe a massive vector, one has
to find an appropriate unfolded field redefinition. But first we have
to learn how an unfolded system for a massive vector field looks like.

Using Fierz\textendash Pauli formulation, a massive spin-1 field corresponds
to the system
\begin{equation}
(\square+m^{2})b_{\alpha\dot{\alpha}}(x)=0,\quad\nabla^{\alpha\dot{\alpha}}b_{\alpha\dot{\alpha}}(x)=0.
\end{equation}

We see that, in order to describe a massive spin-1 field, one can
simply take the equation \eqref{J_united_eq} for $\mathcal{J}$ and
put $\square\mathcal{J}=-m^{2}\mathcal{J}$ \cite{unf9}. Then an
unfolded equation for a massive vector field is
\begin{equation}
\mathrm{d}_{L}\mathcal{J}-\frac{1}{(N+1)(\bar{N}+1)}\left\{ \nu e\partial\bar{\partial}\mathcal{J}-(\nu+2)ey\bar{y}m^{2}\mathcal{J}-e\bar{y}\partial(\mathrm{J}^{+}+2m^{2}\mathcal{J})-ey\bar{\partial}(\mathrm{J}^{-}+2m^{2}\mathcal{J})\right\} =0.\label{massive_vector_eq}
\end{equation}

Now let us analyze \eqref{higgs_Theta_eq} more closely. A ground
equation at $Y=0$ is the only equation that the term with $A$ contributes
to,
\begin{equation}
\mathrm{d}_{L}\Theta(Y=0)-q\phi_{0}A-e\partial\bar{\partial}\Theta(Y=0)=0.\label{eq_Theta_Y_0}
\end{equation}
Introducing $Y$-expansion of $\Theta$ as
\begin{equation}
\Theta(Y|x)=\sum_{n=0}^{\infty}\theta_{\alpha(n),\dot{\alpha}(n)}(x)y^{\alpha_{1}}...y^{\alpha_{n}}\bar{y}^{\dot{\alpha}_{1}}...\bar{y}^{\dot{\alpha}_{n}},\label{Theta_Y_expansion}
\end{equation}
the equation \eqref{eq_Theta_Y_0} gives
\begin{equation}
\theta_{\alpha\dot{\alpha}}=\nabla_{\alpha\dot{\alpha}}\theta-q\phi_{0}A_{\alpha\dot{\alpha}}.\label{theta_A}
\end{equation}
Using the linearized gauge symmetry \eqref{Phi_transform}, one can
gauge away $\theta(x)$, then $\eqref{theta_A}$ identifies $\theta_{\alpha\dot{\alpha}}$
with $A_{\alpha\dot{\alpha}}$ (in this gauge). Note that $\theta_{\alpha\dot{\alpha}}$
itself is gauge-invariant, as well as all higher descendants.

Then we introduce a new unfolded gauge-invariant master-field

\begin{equation}
B:=-\frac{1}{q\phi_{0}}e^{\nabla y\bar{y}}\theta_{\alpha\dot{\alpha}}y^{\alpha}\bar{y}^{\dot{\alpha}}=-\frac{N}{q\phi_{0}}\Theta.\label{B_def}
\end{equation}

Substituting this back into \eqref{higgs_F_eq} and \eqref{higgs_Theta_eq}
yields 
\begin{eqnarray}
 &  & \mathrm{d}_{L}F-\frac{1}{(N+1)(\bar{N}+1)}\left\{ \nu e\partial\bar{\partial}F-(\nu+2)m_{V}^{2}ey\bar{y}F-2m_{V}^{2}ey\bar{\partial}B\right\} =0,\label{higgs_F_final_eq}\\
 &  & \mathrm{d}_{L}B-\frac{1}{(N+1)(\bar{N}+1)}\left\{ \nu e\partial\bar{\partial}B-(\nu+2)m_{V}^{2}ey\bar{y}F-e\bar{y}\partial F-ey\bar{\partial}\bar{F}\right\} =0,\label{higgs_B_finaL-eq}
\end{eqnarray}
where
\begin{equation}
m_{V}^{2}=2q^{2}\phi_{0}^{2}.
\end{equation}
We see that equations \eqref{higgs_F_final_eq}-\eqref{higgs_B_finaL-eq}
(plus a conjugate equation for $\bar{F}$) are indeed equivalent to
the unfolded system \eqref{massive_vector_eq} for a massive vector.
Equations for two helicities of the massless photon receive linear
corrections that couple them to the equation for the imaginary part
of the scalar field, which after performing unfolded field redefinition
\eqref{B_def} becomes an equation for the longitudinal polarization
of the massive vector. Note that this field redefinition is non-invertible,
because the Euler operator $N$ amputates $Y$-independent primary
component $\theta(x)$, thus drastically changing the structure of
the unfolded module: now the module describes a vector instead of
a scalar. This is not inconsistent, because $U(1)$ gauge symmetry
turns $\theta(x)$ to the pure gauge over higgsed $\phi_{0}$-vacuum,
while all other fields in the module are gauge-invariant. Finally,
the equation \eqref{dA_QED} now becomes just a particular consequence
of the equations \eqref{theta_A}, \eqref{B_def} \eqref{higgs_B_finaL-eq}.

Thus, we revealed the picture of the Higgs mechanism within the framework
of the unfolded dynamics approach. It includes a changeover of unfolded
modules caused by the gauge symmetry, whose action is modified by
a nontrivial vacuum. To improve our understanding of unfolded spontaneous
symmetry breaking is especially important from the point of view of
HS symmetry breaking in HS gravity, presumably resulting in the emergence
of string theory as a symmetry-broken phase.

\section{Conclusion\label{SEC_conclusions}}

In the paper, we put forward a novel method of unfolding field theories,
based on postulating a specific form of an unfolded field and the
subsequent search for the corresponding unfolded equation as an identity
that this field satisfies. We successfully apply this method to the
problem of unfolding scalar electrodynamics, where unfolding map is
strongly nonlinear due to the presence of gauge interaction. This
way we end up with a system of nonlinear unfolded equations for which
we have a manifest solution, which is a representation for unfolded
field that we started with. A curious feature of this system of unfolded
equations is that it contains cubic terms made up of scalar fields
solely. These terms might look like charged-current interactions,
but in fact they do not correspond to any real vertices and represent
just artifacts of the nonlinearity of unfolding that underlies the
system.

Considering a particular form of a scalar potential, we are able to
study the spontaneous symmetry breaking in this system. We identify
an appropriate non-invertible unfolded field redefinition that allows
us to reproduce the correct spectrum of the symmetry-broken phase
and study the concomitant deformation of unfolded modules. This is
interesting in the context of recent research on spontaneous symmetry
breaking in HS gravity \cite{HSbreak}.

It would be interesting to apply the proposed method of unfolding
to more complicated theories like e.g. Yang-Mills theory or gravity.
Another topical question is to study the problem of integrability
of unfolded systems: the presented nonlinear unfolded system has a
manifest solution that reconstructs the dependence on auxiliary spinor
variables of a given solution of the space-time equation. The problem
is to develop an algorithm that would allow one to reconstruct the
space-time dependence from the spinorial one, i.e. to generate solutions
to classical e.o.m. starting from Cauchy data.

\section*{Acknowledgments}

The author is grateful to S.M. Ailuro for pointing out a typo and
to the Referee for the suggestion to consider the possibility of encoding
nonlinear deformations by imposing constraints on a linear unfolded
system.

\end{document}